# Ordered InAs QDs using prepatterned substrates by monolithically integrated porous alumina

Pablo Alonso-González, María S. Martín-González, Javier Martín-Sánchez, Yolanda González and Luisa González

*Instituto de Microelectrónica de Madrid (CNM, CSIC), Isaac Newton 8, 28760, Tres Cantos, Madrid, Spain*

Abstract

In this work, we explore a method for obtaining site-controlled InAs quantum dots (QDs) on large areas of GaAs (0 0 1) pre-patterned surface. The patterning of the substrate is obtained by using a monolithically integrated nano-channel alumina (NCA) mask and transferring its self-ordering to the underlying GaAs substrate by continuing the anodization process once the GaAs surface is reached. After patterning, the GaAs substrate follows a low temperature process for surface preparation before epitaxial growth for QD formation. As a final result, we observe that the nanoholes act as preferential nucleation sites for InAs QD formation, with a filling factor close to unity, while the QD formation on the surface region between the pattern holes is completely suppressed.

1. Introduction

Semiconductor quantum dots (QDs) have attracted much attention during the last decade due to their special opto-electronic properties [1]. As a result, a variety of novel devices have been developed and much more predicted [2]. Moreover, its particular nature of fully quantized electronic states together with the possibility of manipulation as artificial atoms makes them perfect candidates in order to study new phenomena predicted by theoretical quantum physics studies. In order to obtain an actual advantage of their properties, it is mandatory to develop technological processes that allow to fabricate QD with control in size, shape and position. Thus, the natural randomly QD nucleation of self-assembled Stranski–Krastanov (SK) growth [3] has to be overcome, while keeping the main characteristics of this spontaneous process: QD formation without defects introduced by fabrication processes.

The use of pre-patterned substrates is a quite wide-spread strategy. In this direction, highly ordered arrays of QD have been obtained using different lithographic approaches [4], [5], [6]. The responsible mechanisms involved in QD selective formation on patterned substrates are not completely clear up to now. Different proposals have been reported in the literature mainly related to the presence of highly reactive stepped and faceted surfaces and to the preferential aggregation of atoms due to the gradient of chemical potential in the patterned surfaces [7], [8], [9], [10].

One common approach for obtaining patterned surfaces is to transfer the pattern from a mask attached. In particular, self-ordered nano-channel alumina (NCA) mask fabricated by electrochemical anodization [11] of an aluminium film, has been previously demonstrated for fabrication of ordered 2D arrays of nanostructures [12], [13], [14], [15], [16].

Our particular approach consists of using an epitaxial crystalline aluminium layer grown on GaAs (0 0 1) substrates by molecular beam epitaxy (MBE) as starting point for further porous alumina fabrication. This layer is subsequently anodized in an electrochemical cell. By this process we obtain a GaAs substrate with a monolithically integrated nanoporous alumina mask. Continuing the anodization process a few nanometres through the GaAs substrate and removing the NCA mask, we finally obtain a large area of ordered nanoholes at the GaAs substrate.

The advantage of this procedure is that the technological processes necessary to fix the alumina mask on top of GaAs are suppressed and consequently the related GaAs surface contamination avoided. Our results show that after InAs deposition, QD are only formed inside the holes and not randomly across the surface, demonstrating that the pattern holes obtained by this approach act as preferential nucleation sites for InAs.

2. Experimental procedure

The experimental procedure is schematically summarized in Fig. 1. The process started growing a 500 nm thick Si-doped ($1\times10^{18}$ cm$^{-3}$) GaAs buffer layer on a GaAs (0 0 1) Si-doped substrate by MBE at a growth temperature of $T_s$=580 °C (Fig. 1(a)). After this buffer layer growth, the samples were cooled down, therefore, from 580 °C to RT, with a chamber base pressure of $P$=6×10$^{-9}$ mbar. During this process, the GaAs surface always maintained a well-defined As-rich GaAs (2×4) surface reconstruction. An epitaxial Al layer was grown on this RT GaAs (2×4) surface. The first 7 nm thick Al layer was grown at low growth rate in order to observe in detail this first stages of Al layer growth. In particular, a growth rate corresponding to 0.1 monolayer per second (ML/s) of AlAs on GaAs (0 0 1) was used, as previously measured from reflection high energy electron diffraction (RHEED) oscillations. The rest of the Al layer was grown at a growth rate corresponding to $r_g$ (AlAs)=1.68 ML/s (Fig. 1(b)).

Once the Al/GaAs sample was taken out from the MBE chamber and carried into the electro-chemical set-up, the NCA mask fabrication process started. This was, firstly, synthesis of ordered nanoporous alumina by electrochemical anodization of the Al layer and finally, transfer of the self-organized pattern of the mask to the GaAs substrate.

A two-step Al anodization process [17] was carried out in a two electrode cell using a regulated DC power supply. The sample was anodized in a 0.3 M oxalic acid at 40 V and temperature $T$=6 °C while the electrolyte was mechanically stirred. Platinum gauze was used as a counter electrode. The first anodization process was stopped after 466 nm of alumina formation. At this moment, the alumina layer was selectively removed by immersing in 4 vol% CrO$_3$+10 vol% H$_3$PO$_4$ for 5 min at 70 °C. The periodic concave features shown at the Al surface after removal of the alumina formed during the first anodizing step (Fig. 1(c)), were used as seeds for further formation of ordered pores at the second anodization step. This second process was made under the same conditions as in the first anodization step. The anodization was stopped once the GaAs surface was reached and, therefore, after the total anodization of the Al was achieved. This critical point was recognizable by means of a colour change of sample surface and a sharp increase of the current intensity in the chronoamperometry, as shown on Fig. 2. Maintaining this situation for approximately 60 s, the GaAs surface under the alumina nanoholes was attacked and thus the alumina self-ordered pattern was transferred to the GaAs substrate (Fig. 1(d)). Finally, the porous alumina fabricated was selectively etched away by a few seconds dip in a 49% dissolution of fluorhidric acid (Fig. 1(e)).

As a result of the above described process, we obtain a pattern in the GaAs substrate. Next experimental step was the growth of InAs QDs on the patterned GaAs surface. For that, an optimal epitaxial growth process has to be performed at a low enough temperature in order to inhibit pattern smoothing.

The sample was introduced again in the MBE chamber together with a non-patterned epitaxial GaAs as a reference. Before the deposition of InAs for QD formation, native oxide and other contaminants of the surface were removed by exposing the GaAs surface of the substrates (reference and patterned) pattern for 5 min to an atomic hydrogen flux [18] with a hydrogen pressure of $P$(H$_2$)=10$^{-5}$ mbar at a substrate temperature $T_s$=450 °C.

After the oxide removal, a 6.8 nm thick GaAs buffer layer was grown at low temperature ($T_s$=490 °C) by atomic layer molecular beam epitaxy (ALMBE) technique [19] (Fig. 1(d)).

The formation of InAs QD was carried out by growing InAs up to critical thickness ($\theta_c$=1.7 ML), as observed by a 2D–3D change in the RHEED diagram of the reference sample (without pattern) at $T_s$=490 °C. For InAs deposition, a growth sequence consisting of 0.1 ML of InAs deposition at a growth rate of 0.07 ML/s was used, followed by a pause of 2 s under $As_2$ flux.

The whole growth process was monitored in situ by RHEED. Scanning electron microscopy (SEM), atomic force microscopy (AFM) and X-ray diffraction characterization techniques were employed for controlling the above exposed fabrication process.

3. Results and discussion

Fig. 3 shows a $\theta/2\theta$ X-ray diffractogram from the Al epitaxial layer. Besides the GaAs substrate-related reflections, we observe two other peaks that correspond to Al(1 1 0) and Al(0 0 1) crystal orientations. Considering the structural factors of these two reflections, the amount of material showing Al(0 0 1) orientation in our Al epitaxial layer is about 300 times lower than Al(1 1 0) structure. These results are consistent with the RHEED diagram observed at the onset of the Al growth, which also showed a mixture of two crystalline orientations. In fact, at the onset of the Al layer growth we observed a spotty pattern consisting of the superposition of two different lattice parameters that can be related to the two crystalline orientations mentioned above. As the growth continued, streaks appeared in the RHEED diagram along the GaAs [1 1 $\bar{0}$] direction, with the same separation as to that shown by a $2x$ periodicity in a GaAs (0 0 1) lattice (4 Å). Along the GaAs [1 1 0] direction, the Al epitaxial layer RHEED diagram showed also streaks, but with a larger separation, corresponding to a lattice parameter of 2.86 Å. These results imply that the Al layer has a (1 1 0)R orientation [20]. This RHEED pattern was observed after an amount of Al deposited equivalent to 110 nm of AlAs on GaAs (0 0 1). According to the X-ray results, this thickness would correspond to a 40 nm thick Al(1 1 0)R layer.

Thus, RHEED and X-ray diffraction results indicate the presence of a preferential Al(1 1 0)R orientation, although in a region close to the GaAs interface there is a coexistence of the two orientations: Al(0 0 1) and Al(1 1 0)R. Using these results, we estimate a total thickness of 885 nm in the Al layer grown.

The Al surface topography was studied by AFM (not shown). It showed a directional roughness along the [1 1 $\bar{0}$] direction of the GaAs substrate with an rms=1.51 nm. This topography has a certain influence on the Al layer anodization process, as the alumina nanopores formed after the first anodization step are inside the valleys aligned along [1 1 $\bar{0}$] direction (Fig. 4(a)).

Together with the alignment of the nanopores along the main roughness direction of the Al layer, we also observe a local incipient hexagonal periodicity. This is a surprising result, taking into account that only a thickness of 466 nm of Al has been attacked instead of the tens of microns that are normally anodized during the first anodization step when Al bulk is used as source material for fabrication of NCA templates [21].

In order to improve the final periodicity of the pattern, a second anodization step was carried out [17]. Fig. 4(b) shows SEM images of nanoporous alumina pattern obtained after this second anodization step. We observe a clear improvement in the order of the nanoporous alumina membrane obtained in this second step. In this case, the holes obtained show an interpore distance of about 100 nm that perfectly corresponds to the use of 0.3 M oxalic acid as electrolyte in the electro-chemical process [22].

Fig. 5 shows AFM images of both the alumina surface after a double step process (Fig. 5(a) and (c)) and the corresponding GaAs surface once the nanoporous alumina layer was selectively removed (Fig. 5(b) and (d)).

Although the alumina pattern seems to be quite homogeneous across large areas of the surface, the homogeneity of the nanoholes distribution that results at the GaAs surface is much smaller. We find flat areas without holes, where the GaAs anodization did not take place, coexisting with hollows where it seems that closely formed nanoholes have collapsed. These differences are related to the formation process of alumina nanoholes during the aluminium anodization: first of all, there is a time delay in the nucleation among the different pores and, secondly, some of the pores grow more perpendicular to the GaAs surface than others. This means that the total thickness of the porous alumina film is not the same for both types of pores. For these reasons, at a certain time, some of the pores reach the GaAs surface and start to widen while others have not yet reached the GaAs surface giving the morphology shown in Fig. 5(b) and (d).

In the regions where there has been an efficient pattern transfer (Fig. 5(d)), the density of holes of the GaAs substrate, $5\times10^9$ cm$^{-2}$, coincides with the alumina nanoholes density (Fig. 5(c)). The corresponding depth of the patterned holes on the GaAs substrate range from 10 to 20 nm.

It is also shown (profile on bottom of Fig. 5(d)), that certain facets are originated inside the holes during the GaAs anodization process.

The patterned GaAs substrates, together with a piece of unpatterned GaAs reference substrate, were introduced in the MBE reactor for the growth of InAs QD. After oxide removal by atomic hydrogen (as described in the "Experimental procedure" section) a 6.8 nm thick GaAs buffer layer was grown at low temperature by ALMBE. During growth of this thin buffer layer, the RHEED pattern observed on the flat reference sample corresponded to a 2D growth process. In the case of the patterned surface, the AFM images (not shown) of the surface after buffer layer growth showed a total preservation of the pattern and smooth regions between holes.

Finally, we deposited 1.7 ML of InAs, the critical thickness for QD formation as we observed by 2D–3D transition in the RHEED pattern of reference sample. AFM images of the patterned surface after InAs deposition (Fig. 6) show that the nanoholes pattern act as preferential nucleation centres for InAs with total absence of QD formed outside them. Thus, apart from the patterned nanoholes, there are not any other preferential nucleation centres for InAs QD formation.

Looking at the AFM image (Fig. 6) it is clearly observed that there is a certain distribution in QD size. Considering the large non-uniformities in the patterned motives shown in Fig. 5(d) (we observe a SD of 26% centred at a mean diameter of 89 nm), we find quite plausible to establish a correlation between the dispersion in the nanoholes size and the resulting distribution of QD size. In this respect, we could fix an upper limit for QD size uniformity corresponding to the nanoholes size dispersion. This value is larger than that obtained in the QD distribution of the reference sample (SD of 12% at a mean diameter of 50 nm).

These results show that although the whole process that demonstrate selective nucleation of QD, the characteristics of the patterned surface must be improved in order to achieve a highly uniform distribution of QD formed at specific sites. In this situation, we think that the non-uniform alumina hole depth hinders the simultaneous transfer of the pattern across the surface, with a final result of a non-homogeneous distribution of the nanoholes. Fortunately, these are not intrinsic problems associated to the porous anodic alumina templates, but can be overcome by following different approaches [23].

4. Conclusions

A large area process for obtaining site-controlled InAs QD in pre-patterned GaAs (0 0 1) surface has been studied. The patterned GaAs surface has been produced by transferring the pattern of an electrochemically anodized nanohole alumina layer fabricated on a single crystal Al layer directly grown on the GaAs substrate by MBE.

Our results demonstrate that the nanohole alumina membrane geometry can be directly transferred to the GaAs substrate. The monolithically integrated Al layer act as a good mask that preserves the cleanliness and smoothness of the GaAs (0 0 1) substrate during the processes followed for GaAs patterning. The subsequent process of surface preparation for epitaxial growth preserves the patterning and produces clean and smooth surface regions between holes. As a consequence, InAs QD were only formed inside the nanoholes, showing that nanoholes obtained by GaAs anodization act as preferential sites for InAs growth.

Acknowledgements

This work was financed by Spanish MEC under NANOSELF II project (TEC2005-05781-C03-01), NANOCOMIC project (CAM S 0505ESP 0200) and by the SANDIE Network of excellence (Contract no. NMP4-CT-2004-500101 group TEP-0120). M.S. Martín-González thanks to the Ramón y Cajal programme. P. Alonso-González and J. Martín-Sánchez thanks to European Community for a fellowship under the Programme I3P-CSIC.

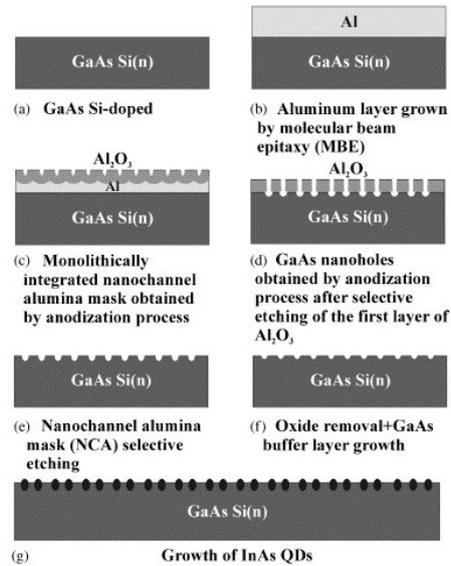

Fig. 1. Scheme of the fabrication process of 2D array of ordered InAs/GaAs (0 0 1) quantum dots developed in this work.

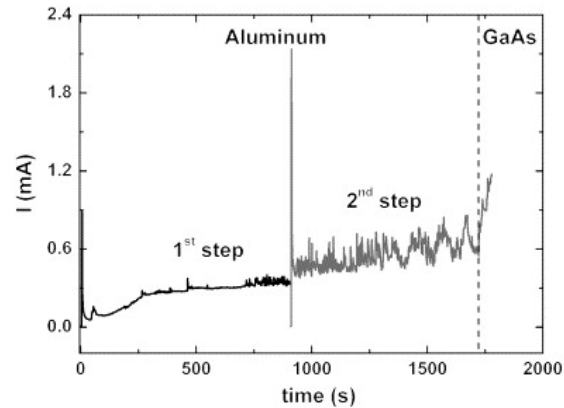

Fig. 2. Cronoamperometry showing the differences in current density between the first and second anodization step. Between both steps a selective dissolution of the porous alumina is performed. A drastic slope change in the intensity is observed when the semiconductor surface is reached (mark with a dotted vertical line).

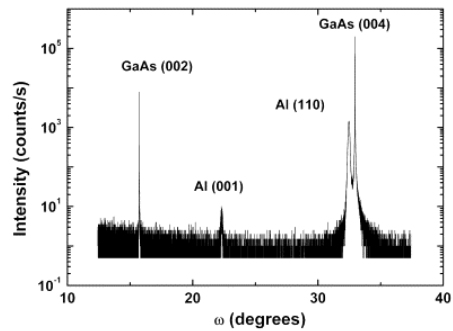

Fig. 3. $\theta/2\theta$ double crystal X-ray diffractogram of Al epitaxial layer grown on GaAs (0 0 1) substrate, showing diffraction peaks corresponding to (1 1 0) and (0 0 1) Al crystal orientations.

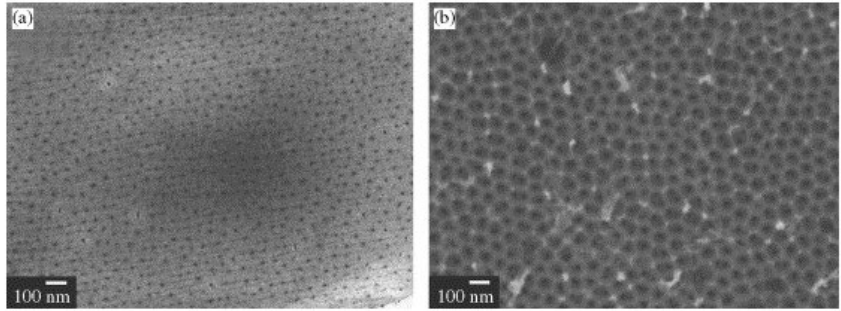

Fig. 4. SEM images of nanoporous alumina layer surface fabricated on Al/GaAs (0 0 1) at different stages of the fabrication process: (a) after first step and (b) after second step of the anodization process. Observe the improvement in ordering in the alumina layer obtained by a two-step process.

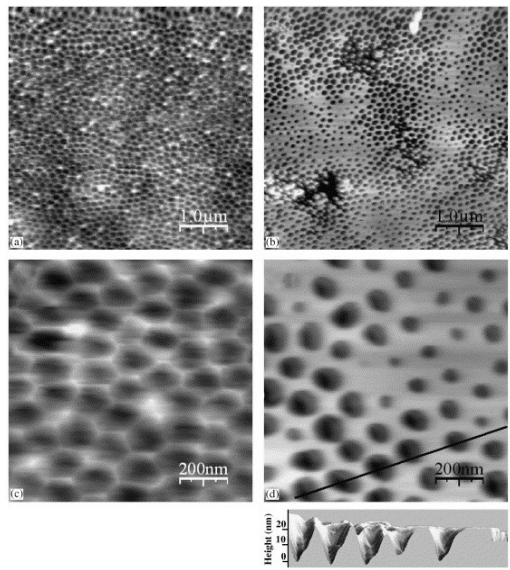

Fig. 5. AFM images of alumina surface after the second step of the anodization process (a,c) and patterned GaAs surface resulting after alumina removal (b,d). A profile along the line drawn in (d) indicates faceting at the GaAs nanoholes (bottom part of (d)).

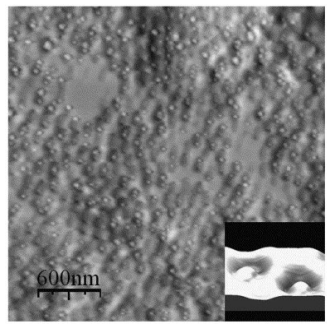

Fig. 6. AFM surface image obtained after the deposition of 1.7 monolayers of InAs on GaAs patterned substrate. A surface topography derivative is shown to emphasise the morphology features. Observe that most of the GaAs nanoholes are occupied by an InAs quantum dot (QD) and no QD are outside the holes. At the bottom inset is a 3D topographic image of the QD.